\documentclass[fleqn]{article}
\usepackage{enumerate,graphicx,natbib,amsfonts,amsbsy}
\usepackage[mathscr]{eucal}
\bibpunct{(}{)}{;}{a}{,}{,~}
\DeclareMathAlphabet\EuRscript{U}{eur}{m}{n}
\SetMathAlphabet\EuRscript{bold}{U}{eur}{b}{n}
\newcommand{\cur}{\EuRscript}
\DeclareSymbolFont{AMSeur}{U}{eur}{m}{n}
\DeclareSymbolFont{AMSeub}{U}{eur}{b}{n}
\DeclareMathSymbol{\curdelta}{\mathbin}{AMSeur}{"0E}
\DeclareMathSymbol{\curpi}{\mathbin}{AMSeur}{"19}
\DeclareMathSymbol{\gtrless}{\mathrel}{AMSa}{"3F}
\DeclareMathSymbol{\lessgtr}{\mathrel}{AMSa}{"37}
\DeclareMathSymbol{\approxeq}{\mathrel}{AMSb}{"75}
\newcommand{\bmf}[1]{\boldsymbol{#1}}
\newcommand{\eus}[1]{\mathscr{#1}}

\newcommand{\BbbR}{\mathbb{R}}
\newcommand{\BbbZ}{\mathbb{Z}}
\newcommand{\fI}{\mathfrak{I}}

\newcommand{\bx}{\bmf{x}}
\newcommand{\bq}{\bmf{q}}
\newcommand{\bp}{\bmf{p}}
\newcommand{\eusF}{\,\eus{F}\,}
\newcommand{\eusG}{\,\eus{G}\,}
\newcommand{\curm}{\cur{m}}
\newcommand{\mcB}{{\mathcal{B}}}
\newcommand{\mcL}{{\mathcal{L}}}
\newcommand{\tA}{{{\mbox{\tiny A}}}}
\newcommand{\tB}{{{\mbox{\tiny B}}}}
\newcommand{\tC}{{{\mbox{\tiny C}}}}
\newcommand{\tG}{{{\mbox{\tiny G}}}}

\newcommand{\tE}{{{\mbox{\tiny E}}}}
\newcommand{\tD}{{{\mbox{\tiny D}}}}
\newcommand{\tT}{{{\mbox{\tiny T}}}}
\newcommand{\mx}{{{\mbox{\tiny Max}}}}

\newcommand{\tmN}{{{\mbox{\tiny$N$}}}}
\newcommand{\prb}{{\sf Prob}}
\newcommand{\prbs}{{\sf p}}
\newcommand{\ttime}{{\sf T}}
\newcommand{\dd}{{\mathrm{d}}}
\def\pairsep{\hspace{1cm}}
\def\tripsep{\hspace{0.7cm}}
\def\tfrac#1#2{{\textstyle\frac{#1}{#2}}}
\def\ssfrac#1#2{\mbox{\large$\frac{#1}{#2}$}}
\title{Boltzmann and Gibbs: An Attempted Reconciliation\footnote{This is
a modified version of the conference contribution presented under
the title of ``Is Equilibrium a Useful Concept in Statistical
Mechanics?''.\break To appear in Stud. Hist. Phil. Mod. Phys. June
2005.}}
\author{ D.\ A.\ Lavis\\
Department of Mathematics, King's College,\\
Strand, London WC2R 2LS, U.K.\\
\small{Email:David.Lavis@kcl.ac.uk}
}

\date{}

\begin{document}

\maketitle
\begin{abstract}
There are three levels of description in classical statistical mechanics,
the microscopic/dynamic, the macroscopic/statistical and the thermodynamic.
At one end there is a well-used concept of equilibrium in
thermodynamics and at the other dynamic equilibrium
does not exist in measure-preserving reversible dynamic systems.
Statistical mechanics attempts to situate equilibrium at the macroscopic level in
the Boltzmann approach and at the statistical level in the Gibbs approach. The
aim of this work is to propose a reconciliation between these approaches and
to do so we need to reconsider the concept of equilibrium. Our proposal is
that the binary property of the system being or not being in equilibrium
is replaced by a continuous property of {\em commonness}.\break\break
\end{abstract}

%----------------------------------------------
\section{Introduction}\label{intro}
One of the fundamental problems in the foundations of statistical
mechanics is to give an explanation as to why `equilibrium'
statistical mechanics is so successful. That is to say, why the use
of the standard Gibbsian methods
`reproduces' thermodynamic results. One offered explanation for
this is the standard ergodic approach. As to whether this gives an
acceptable explanation \citet{vanL1} offers the impression
``that the {\em communis opinio} in the physics literature is that
it does; in the philosophy literature that it doesn't''.
My impression agrees with hers. However, there is a further
twist to the story. When confronted with the question of what is
`actually going on' in a gas of particles (say) when it is in equilibrium,
or when it is coming to equilibrium, many physicists
are quite prepared to desert the Gibbsian approach entirely and to
embrace a Boltzmannian view \citep{rue2,leb1,bric,gold1}.
In particular according to \citeauthor{gold1} (ibid):
\begin{quote}
``Ludwig Boltzmann explained how irreversible macroscopic laws $\ldots$
originate in the time-reversible laws of macroscopic physics. Boltzmann's
analysis $\ldots$ is basically correct. The most famous criticisms of Boltzmann's
later work on the subject have little merit. Most twentieth century innovations
-- such as the identification of the state of a physical system with a probability
distribution $\rho$ on phase space, of its thermodynamic entropy with the Gibbs
entropy of $\rho$, and the invocation of the notions of ergodicity and mixing for
the justification of statistical mechanics -- are thoroughly misguided'' (p.\ 39).
\end{quote}
and \citeauthor{leb1} (ibid):
\begin{quote}
``Having results for
typical microstates rather than averages is not just a
mathematical nicety but at the heart of understanding the
microscopic origin of observed macroscopic behaviour. We
neither have nor do we need ensembles $\ldots$.
What we do need and can expect is typical behaviour'' (p.\ 38).
\end{quote}
These assertions are reinforced by the opinion of
\citeauthor{rue2}\footnote{I think Ruelle rather overstates the case. In particular the rational
subjectivist approach of \citet{jay}
and the interventionist approach most recently argued for by \citet{rid&red}
and \citet{rid} would find some favour.} (ibid)
that the Bolzmannian approach
\begin{quote}
``is now generally accepted by physicists.
$\ldots$ There are some dissenting voices, such as that of Ilya
Prigogine, but the disagreement is based on philosophical prejudice rather
than physical evidence'' (p.\ 113).
\end{quote}
However, most work in equilibrium statistical mechanics uses the
tools developed by Gibbs. Given a particular thermodynamic setup and
microscopic model the appropriate probability distribution
(microcanonical, canonical, grand-canonical etc.) is chosen. The
entropy is taken to be that of Gibbs and the holy grail for any
investigation is an analytic form for the partition function; the
notable successes being the solution of the zero-field
two-dimensional Ising model by \citet{onsager}, of the six-vertex
model in 1967 by Lieb \citep[see,][]{lieb&wu} and of the eight-vertex
model in 1972 by Baxter \citep[see,][]{bax}. There have been many
attempts to extend the Gibbs approach to non-equilibrium.
As indicated above in the
quote from \citeauthor{rue2}, the most developed programme for
doing this is that of the Brussels--Austin School of
the late Count Ilya Prigogine.\footnote{Although the fundamentals of
the work of this school have remained unchanged since their
inception in the early sixties, the approach has evolved substantially,
leading to a great wealth of publications. For a comprehensive review
see \citet{bish3}.} However,
many remain unconvinced of either the actual or potential success of
this enterprize. There would seem to be the need to explore the
possibility of holding, with \citeauthor{leb1}, \citeauthor{gold1},
\citeauthor{rue2} et al., to the conviction that Boltzmann got it
right about the approach to equilibrium, whilst at the same time
with a good conscience continuing to use the standard distributions
of Gibbs for everyday equilibrium calculations. We shall attempt to
take some steps along this path. In order to do so we need to
resolve in some way three questions, to which the current versions
of the Gibbs and Boltzmann approaches offer apparently
irreconcilable answers.
\begin{enumerate}[(i)]
\item What is meant by equilibrium?
\item What is statistical mechanical entropy?
\item What is the object of study?
\end{enumerate}
The attempt to produce conciliatory answers  to (i) and (ii) will occupy most of this paper. However,
we must at the outset deal with (iii). Ensembles are an intrinsic feature of the approach of
\citeauthor{prig1} for whom ``it is at the level of ensembles
that temporal evolution can be predicted'' \citep[p.\ 8]{prig1} and for
 \citet{mac1} a ``thermodynamic system
{\em is} [my italics] a system that has, at any given time, states
distributed throughout phase space $\ldots$, and the distribution of
these states is characterized by a density $\ldots$''(p.\ 984). The density
referred to is the ensemble density and thus the ensemble
becomes the way that a thermodynamic system is {\em defined}.
In contrast to this we shall follow
the view of \citeauthor{leb1} given above that the object of study in statistical mechanics
is a {\em single system}. All talk of `ensembles' is taken to be just a way of giving
a relative frequency flavour
to the probabilities of events occurring in that system.

In Sec.\ \ref{micros} we define the dynamic microstructure of our system
and introduce two gas models which will be used as examples in the later parts of the paper.
In Sec.\ \ref{approaches}
we describe the different concepts of equilibrium in dynamics, the Boltzmann and Gibbs approaches
to statistical mechanics and thermodynamics  In Sec.\ \ref{bmod}
we discuss the Boltzmann approach with particular reference to the way in which equilibrium is
defined and the Neo-Boltzmannian reliance on the concept of typical behaviour. With respect to the latter
we suggest that the expected typical behaviour can be encapsulated by the term `thermodynamic-like',
which signifies the type of fluctuations to be expected in dependent thermodynamic variables, most
particularly the entropy, and we discuss the conditions on the system needed for thermodynamic-like
behaviour to be typical. At that point it is necessary to consider the usual demarcation made between
the system being or not being in equilibrium and we discuss, with the use of the example
of the baker's gas with a small number of microsystems, the problematic nature of this
property. This leads us to the rather radical suggestion that it would be better to replace
the binary equilibrium property with a continuously changing property which we call {\em commonness}.
In Sec.\ \ref{gibbs} we use the example of a gas expanding in a box when a
partition is removed to describe the role ascribed to the Gibbs approach in the overall picture we
are attempting to develop. Of course, one question which needs to be addressed is the meaning of probability
and this is discussed in Sec.\ \ref{prob}. Sec. \ref{prop} contains our proposals for a model incorporating
both the Boltzmann and Gibbs viewpoints and our conclusions are presented in Sec. \ref{conc}.

%------------------------------------------------------
\section{The Microstructure}\label{micros}
Consider a system that, at time $t$, has
a microstate given by  the vector $\bx(t)$ in the {\em phase space}
$\Gamma$. The time parameter $t$ can be discrete or continuous and the phase space can also be
continuous or discrete.
Some one-to-one autonomous dynamics $\bx\rightarrow \phi_t\,\bx$, ($t\ge 0$)
determines a {\em flow} in $\Gamma$ and the set of points
$\bx(t)=\phi_t \bx(0)$, parameterized by $t\ge 0$,  gives a
{\em trajectory}. The set of mappings $\{\phi_t\}_{t\ge 0}$
is a semi-group.
The system is {\em reversible} if there exists an idempotent operator $\fI$ on the
points of $\Gamma$, such that
$\phi_t\bx=\bx'$ implies that
$\phi_t \fI\,\bx'=\fI\,\bx$. Then
$\phi_{-t}=(\phi_t)^{-1}=\fI\phi_t\fI$ and the set $\{\phi_t\}$ with $t\in {\BbbR}$ or $\BbbZ$ is a
group. Henceforth we shall assume that the system under discussion is reversible.
If $\Gamma$ consists of a finite number of points then $t\in{\BbbZ}$ and the system is
periodic. If $\Gamma$ is continuous then $t\in{\BbbZ}$ or $t\in{\BbbR}$.
In this case, we shall restrict attention to a subset $\Lambda\subseteq\Gamma$
invariant under $\{\phi_t\}$. We denote by $\curm$ a measure on $\Lambda$
such that (a) $\curm(\Lambda)$ is finite, (b) $\curm$ is absolutely continuous with respect to
the Lebesque measure on $\Lambda$, (c) $\curm$ is preserved by the flow; that is
$\curm(\phi_t\gamma)=\curm(\gamma)$, for all $t$ and all
measurable $\gamma\subset\Lambda$. It can now be shown that
the \citet{poin1c} recurrence theorem will apply \citep[p.\ 214]{ott} to flows for
which $\bx(0)\in\Lambda$.

%-----------------------------------------------------
\subsection{Useful Examples}\label{examples}
To clarify the discussion it useful
to resort to computer simulations of simple models. However, most interesting problems in statistical
mechanics concern cooperative systems and, even at equilibrium \citep[see e.g.][]{bax},
there are few of these which can be solved exactly. So, of necessity, useful examples are of
assemblies of non-interacting microsystems and the literature contains discussions
of many `toy models' of this kind.\footnote{A model which has featured prominently in
such discussions is the spin-echo system. For a recent account of this model see \citet{lavis4}
and for simulations for a number of deterministic and stochastic systems see \citet{lavis3a}.}
Here we shall use two.
%----------------------------------------------------
\subsubsection{A Perfect Gas -- Continuous Time and Continuous Space}\label{gas}
\begin{figure}[t]
\begin{center}
\includegraphics[width=90mm,angle=0]{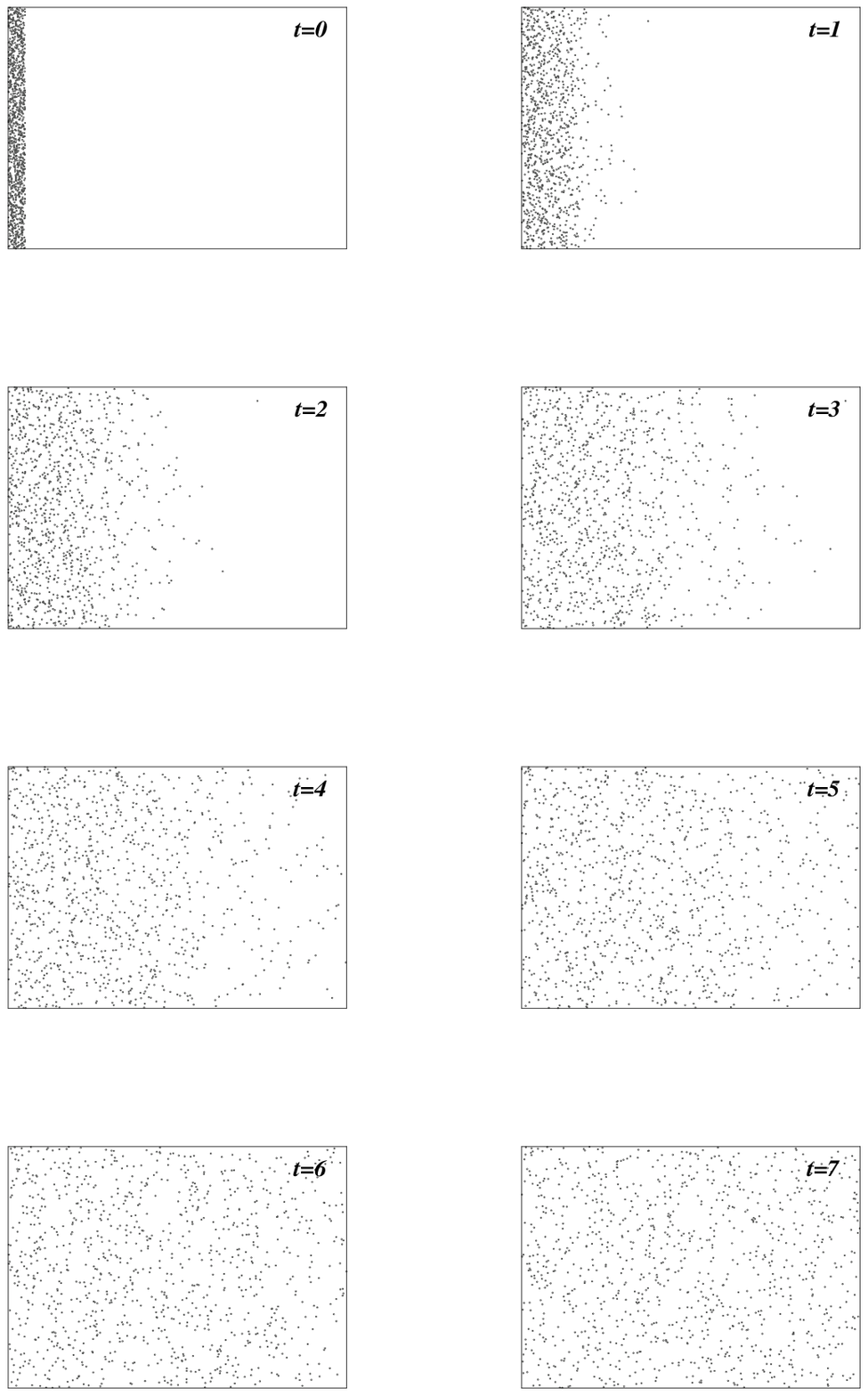}
\end{center}
\caption{A perfect gas of $N=1000$ particles moving horizontally in a box with
elastic walls.}\label{equ-fig1}
\end{figure}

Consider a gas of $N$ particles of unit mass moving in the two-dimensional
box $-L/2\le x\le L/2$,
$-L/2\le y\le L/2$, with elastic walls. We suppose that the initial positions $(x^{(k)}(0),y^{(k)}(0))$,
$k=1,2,\ldots,N$ of the particles are such that all the $y^{(k)}(0)$ are distinct
and the initial velocities are all in the $x$--direction. Then all the particles will
perform one-dimensional oscillations in the $x$--direction at constant speeds. If we denote
the positions and velocities at time $t$ by $(x^{(k)}(t),v^{(k)}(t))$, $k=1,2,\ldots,N$ then
$|v^{(k)}(t)|$ is constant and the system is reversible with $\fI(x^{(k)},v^{(k)})=(x^{(k)},-v^{(k)})$.
In Fig.\ \ref{equ-fig1} we show a typical evolution of the gas from a state where all
the particles are congregated near the left-hand end of the box.
%---------------------------------------------------------------
%-------------------------------------------------------------------------
Calculations for this example are used to illustrate the discussion in Sec.\ \ref{prop}.
%---------------------------------------------------------------------------------
\begin{figure}[h]
\begin{center}
\includegraphics[width=90mm,angle=0]{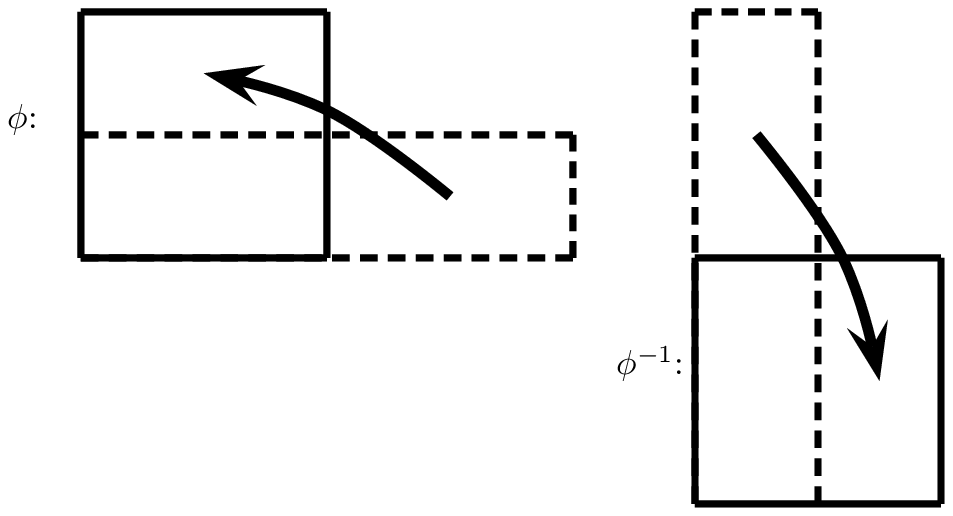}
\end{center}
\caption{The baker's transformation and its
inverse.}\label{equ-fig2}
\end{figure}
%---------------------------------------------------------------------
\begin{figure}[t]
\begin{center}
\includegraphics[width=90mm,angle=0]{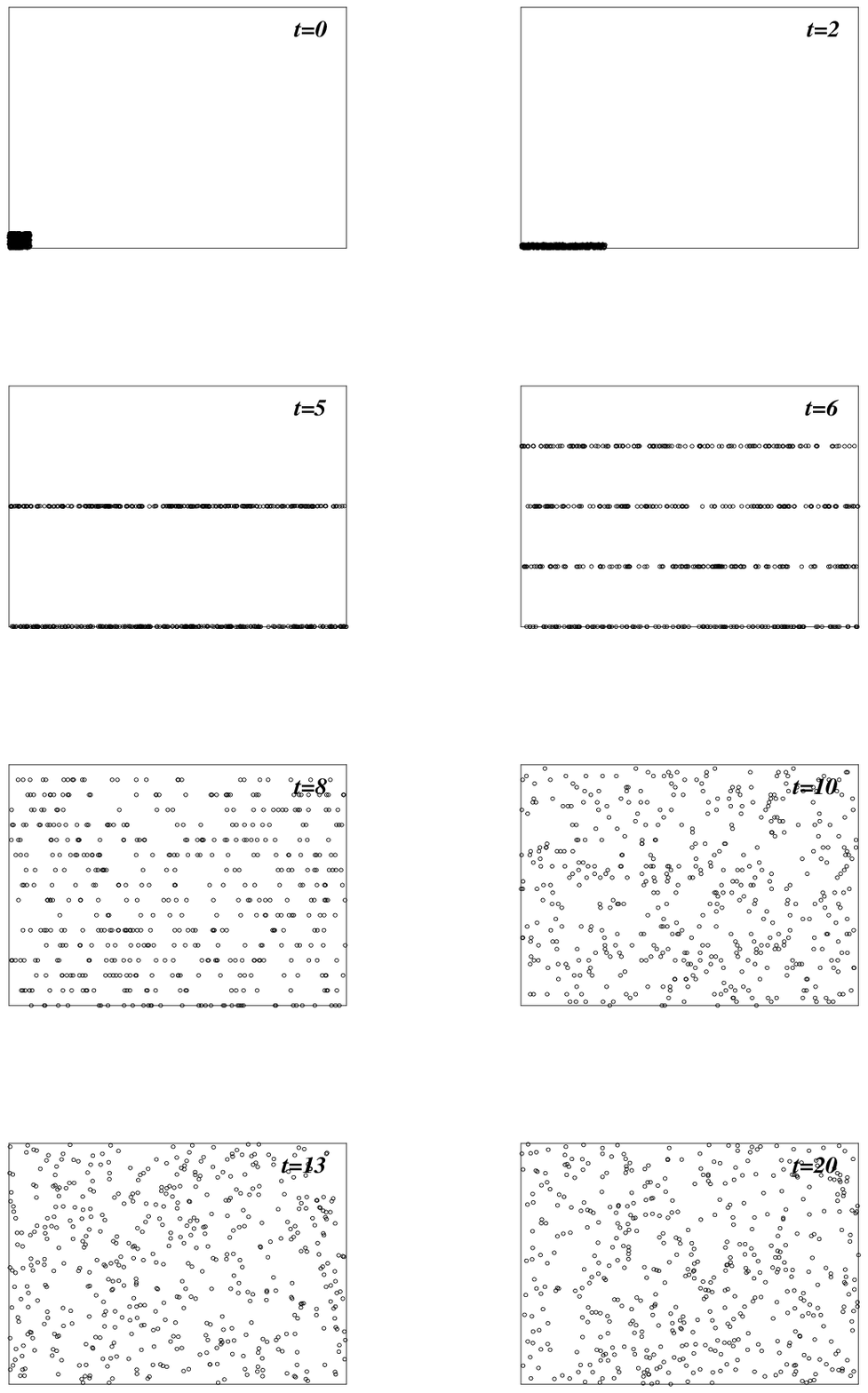}
\end{center}
\caption{A gas of $N=512$ particles moving under the baker's
transformation.}\label{equ-fig3}
\end{figure}
%--------------------------------------------------------------------
\subsubsection{A Baker's Gas - Discrete Time and Continuous Space}\label{baker}
This is the transformation, shown in Fig.\ \ref{equ-fig2}, where
a square of side $L$ is stretched to twice its width and then cut in half
with the right-hand half used to restore the upper half of the
unit square. As the mapping $\phi$ on the cartesian coordinates
$L\times(x,y)$ of the square, it is given by
\begin{equation}
\phi(x,y)= \left\{
\begin{array}{lll}
(2x,\ssfrac{1}{2}y),\quad&{\rm mod}\, 1,\quad&0\le x\le \ssfrac{1}{2},\\[0.3cm]
(2x,\ssfrac{1}{2}[y+1]),&{\rm mod}\, 1,& \ssfrac{1}{2}\le x\le 1.
\end{array}\right.\label{bak1}
\end{equation}
A convenient way of writing this transformation is to express $x$
and $y$ as binary strings:
$x=0\cdot x_1x_2x_3\ldots$ and $y=0\cdot y_1y_2y_3\ldots$ with $x_k,y_k=0,1$.
Then the transformation takes the form
$\phi(0\cdot x_1x_2x_3\ldots,0\cdot y_1y_2y_3\ldots)
 =(0\cdot x_2x_3x_4\ldots,0\cdot x_1y_1y_2\ldots)$, with
$\phi^{-1}(0\cdot x_1x_2x_3\ldots,0\cdot y_1y_2y_3\ldots)= (0\cdot
y_1x_1x_2\ldots,0\cdot y_2y_3y_4\ldots)$.
The mapping is reversible with $\phi^{-1}=\fI\phi\,\fI$
and $\fI(x,y)=(y,x)$.
 It can also be shown \citep[p.\ 54--56]{la&ma}
to be volume-preserving and thus that
the \citet{poin1c} recurrence theorem applies.
We are interested in a baker's gas of $N$ `particles'. In Fig. \ref{equ-fig3}
we show a typical evolution from a state where all the particles are congregated
near the bottom left corner of the box.
%--------------------------------------------------------------------------------------------------
\section{Different Views of Equilibrium}\label{approaches}
The three levels of description in classical statistical mechanics, referred to in the abstract, each bring
with them a definition of equilibrium.
In fact by separating the macroscopic (Boltzmann) and statistical
(Gibbs) approaches one can identify four distinct concepts of equilibrium.

\vspace{0.25cm}

\noindent A \textbf{dynamic system} is in equilibrium at time $t$ if its state is given by a phase point
$\bx(t)$ lying on an attractor.\footnote{For our purpose this is an adequate definition, although the term
is sometimes used only for fixed points, where its use includes repelling (unstable)
and marginal equilibrium points.} However, of course, measure-preserving reversible dynamic systems
do not have attractors.\footnote{This means that the remark by \citet[p.\ 156]{skl} that
in statistical mechanics ``the equilibrium state exists as the `attractor ' to which the dynamics of
non-equilibrium drives systems $\dots$'' must not be taken too literally.}

\vspace{0.25cm}

\noindent To describe the \textbf{Boltzmann approach} we need to introduce some {\em macroscopic variables}.
In general, these will be at the observational
level for the system, but will encapsulate more detail than the thermodynamic
 variables.\footnote{\citet{rid} refers to them as {\em supra-thermodynamic
variables}.} Examples of these would be local variables which quantified spatial inhomogeneities of density
or magnetization in a system (see Sec.\ \ref{bmod}).\footnote{Another simple example, in the context of coin-flipping,
is given by \citet[p.\ 8]{bric1}. In $N$ tosses of a coin a microstate is a particular record
of the $N$ outcomes and a macrostate is identified
by the macrovariable $N_0$ and consists of the set of all microstates for which exactly $N_0$
of the outcomes are heads.}
Suppose that, as in our examples discussed below, the system
consists of $N$ identical microsystems and that we have a set $\Xi$ of
macrovariables. A macrostate $\mu$ is a $\curm$--measurable subset of $\Lambda$.
The set of macrostates $\{\mu\}$ is defined so that:
\begin{enumerate}[(i)]
\item Every $\bx\in\Lambda$ is in exactly one macrostate denoted by $\mu_{\bx}$.
\item Each macrostate corresponds to a unique set of values for $\Xi$.
\item $\mu_{\bx}$ is invariant under all permutations of the microsystems.
\item The phase points $\bx$ and $\fI\bx$ are in macrostates of the same
size.\footnote{This is necessary to ensure that the macrostate structure reflects the
reversibility of the system; which means that the Boltzmann entropy profile
along $\bx \to \phi_t\bx$ is the same as that along $\fI\bx \to \fI\phi_t\bx=\phi_{-t}\fI\bx$.
In fact, of course, in cases, like the perfect gas where macrostates are created by course-graining
the one-microsystem configuration space and reversibility corresponds to changing the sign of the velocity
$\bx$ and $\fI\bx$ are in the same macrostate.}
\end{enumerate}
Given a particular set of macrovariables and macrostates, the {\em
Boltzmann Entropy} is given by
\begin{equation}
S_{\tB}(\bx)=k_{\tB}\ln[\curm(\mu_{\bx})].\label{typ2}
\end{equation}
Equilibrium is defined by reference to the
macrostate, which is uniquely given by a set of values of the
macrovariables. According to \citet[p.\ 179]{bric},\label{briclab} ``by far the
largest volumes correspond to the {\em equilibrium values} of the
macroscopic variables (and this is how `equilibrium' should be
defined)''. This is the standard Boltzmannian definition. The system
is in equilibrium if its phase point is in the macrostate of largest
volume ($\curm$--measure), corresponding to the largest value of
$S_{\tB}$, and this in turn defines the equilibrium values of the
macrovariables.

\vspace{0.25cm}

\noindent The starting point for the \textbf{Gibbs approach} is to suppose the
phase-point $\bx$, in some invariant $\Lambda\subset\Gamma$, is
distributed according to a probability density function\footnote{So
that the probability of the phase point being in a small region
$\curdelta\Lambda$ around a phase point $\bx$ is
$\rho(\bx)\curm(\curdelta\Lambda)$.
 The meaning we give to probability is discussed in more detail
in Sec.\ \ref{prob}.}  $\rho$ which is invariant under the flow (a
solution of Liouville's equation).
Equilibrium is defined as the situation where
the probability density function is not an explicit function of time
and $\rho$ becomes a function of the global constants of motion. The
statistical mechanical `analogues'\footnote{This is the term used by
\citet[chap.\ 14]{gib}.} of thermodynamic quantities are either
fixed external parameters, related to phase functions\footnote{In a
way which will be discussed in Sec.\ \ref{prop}.} or functionals of
$\rho$. In particular the analogue of thermodynamic entropy is the
{\em Gibbs entropy}
\begin{equation}
S_{\tG}[\rho]=-k_{\tB}\int_{\Lambda}
\rho(\bx)\ln[\rho(\bx)]\,\dd\curm. \label{gibbs3}
\end{equation}
From a practical point of view this scheme is very satisfactory. However,
problems arise when an attempt is made to extend it to
non-equilibrium situations, which are now perceived as being
represented by time-dependent solutions of Liouville's equation.
Specifically:
\begin{enumerate}[(i)]
\item When $\rho(\bx)$ is replaced in (\ref{gibbs3}) by any time-dependent solution
$\rho(\bx;t)$ of Liouville's equation, $S_{\tG}[\rho(t)]$ remains
invariant with respect to time.
\item Given an arbitrary initial condition $\rho(\bx;0)$, the evolving solution
$\rho(\bx;t)$ of Liouville's equation will not in general converge to a
time-independent (equilibrium) solution as $t\to\infty$.
\end{enumerate}
The Brussels--Austin programme is, as we have indicated above, an attempt to circumvent these problems.

\vspace{0.25cm}

\noindent In \textbf{classical thermodynamics} the approach to equilibrium is usually described as leaving the system
so that it ``eventually reaches a state in which no further change is perceptible, no matter
how long one waits'' \citep[p.\ 6]{pip}. The equilibrium state is thus the situation in which
(for an isolated system) no perceptible change occurs in any thermodynamic variables. However,
\citet{uff1} has argued that the tendency of isolated systems to approach equilibrium is
not a consequence of the standard laws of thermodynamics. This
has led \citet[p.\ 528]{B&U1}
to formulate a principle which they call the {\em minus first law} that ``an isolated system in an
arbitrary initial state within a finite fixed volume will spontaneously attain a unique state of equilibrium''.
With this addition, thermodynamic equilibrium, for a finite isolated system, can be seen to have
three important properties:
\begin{enumerate}[(a)]
\item It is a binary property; a system either is or is not in equilibrium.
\item A system in equilibrium never evolves away from equilibrium.
\item A system not in equilibrium evolves into equilibrium.
\end{enumerate}
Indeed the quote from Pippard may seem to allow us to add `in a finite time' to (c) or at
the very least that `so that the difference from equilibrium after a finite time is imperceptible'.

\vspace{0.5cm}

\noindent In this description of four distinct types of equilibrium, dynamic equilibrium has been included
for completeness. Although it shares properties (a)--(c) with thermodynamic equilibrium it exists
only for dissipative systems. Statistical mechanical equilibrium for the systems considered here,
for which the dynamics is non-dissipative, cannot be a consequence of, or related to, the system being
in dynamic equilibrium.
The Brussels--Austin extension of the Gibbs programme would also satisfy properties (a)--(c). However,
as will be discussed in Sec.\ \ref{bmod}, the form of equilibrium which features in the Boltzmann approach
does not satisfy (b) and satisfies (c) only in a qualified sense (see Sec. \ref{bmod}).

A classic problem in statistical mechanics (\citealp[chap.\ 9]{skl};\citealp{cal}) is to understand in what
sense thermodynamics can be said to {\em reduce} to statistical mechanics and thus part of the problem of
equilibrium is to relate in some way equilibrium in statistical mechanics and equilibrium in thermodynamics.
However, as we have seen there is no unique statistical mechanical equilibrium. The Boltzmann
and Gibbs approaches have very different concepts. Indeed, with the usual ensemble interpretation of the
probability density function, Gibbsian equilibrium is, unlike Boltzmannian equilibrium, not a property of an
individual system. So for us the problem
is both to effect a reconciliation between Boltzmannian and Gibbsian equilibrium and to relate them to
thermodynamic equilibrium. To do this we propose that equilibrium as a binary property
is replaced by something (which we call commonness)
which encapsulates degrees of `equilibriumness'.

%---------------------------------------------------
\section{The Boltzmann Approach}\label{bmod}
Most examples of the use of this approach consider situations where the macrovariables measure
inhomogeneities in the distribution of microsystems over the one-microsystem phase space. This
space is divided into $p$ cells, labelled $\ell=1,2,\ldots,p$, and, for the phase point $\bx\in \Lambda$,
$N_{\ell}(\bx)$ is the number of microsystems in cell $\ell=1,2,\ldots,p$. Given that the microsystems are
identical, the macrostate $\mu_{\bx}\in \Lambda$ corresponds to all permutations of microsystems in cells with the
same values of the macrovariables. Thus
\begin{equation}
\frac{\curm(\mu_{\bx})}{\curm(\Lambda)}
=\frac{N!}{\prod_{\ell=1}^p\left\{N_{\ell}(\bx)\right\}!}\frac{1}{p^\tmN}.
\label{typ4}
\end{equation}
For the baker's gas $\Lambda$ is the union of $N$ squares of side $L$ giving $\curm(\Lambda)=L^{2\tmN}$
and we divide the one-particle phase space (configuration space in this case) into squares of side
$L/2^m$ to give $p=2^{2m}$.\footnote{The cells are labelled sequentially in rows, so that
cell $(1,1)$ in the bottom left corner is labelled $\ell=1$ and cell $(2^m,2^m)$ in the top
right corner is labelled $\ell=p$.} In the case of the perfect gas $\curm(\Lambda)=L^{\tmN}$ and we take
$p=2^m$.

As indicated above, the entropy in the Boltzmann approach
is the phase function (\ref{typ2}), which will fluctuate with time, the system
being construed as being in equilibrium when entropy has its maximum value
$(S_\tB)_{\mx}$.
The difficulties associated with this definition of
equilibrium are discussed in Sec.\ \ref{ttde}. However, another question arise in
relation to this approach. This is usually
identified as the problem of the `approach to equilibrium'.\footnote{And part of the
discussion is concerned with giving some explanation
as to why the universe seems to have started in a very uncommon state. According to
\citet{gold1} this is the ``hard part'' of the Boltzmann approach. It is
beyond the scope of the present discussion.}
However, it can be extended to the broader problem associated with certain
expectations about the form of the entropy profile and the need to give a dynamic
account which justifies these expectations. In somewhat imprecise language,
we expect that in most cases the entropy will behave in a {\em thermodynamic-like} way.
This we take to mean that:
\begin{quote}
\textsl{The Boltzmann entropy, for the evolving system, is most of the time close to its maximum
value, from which it exhibits frequent small fluctuations and rare large (downward) fluctuations.}\label{thermlk}
\end{quote}
The problem of quantifying and justifying these expectations is discussed in Sec. \ref{typical}.
\subsection{Trying to Define Equilibrium}\label{ttde}
It is clear that $(S_\tB)_{\mx}$ will not necessarily be the value of entropy for
the largest proportion of microstates. Associated with a macrostate $\mu$, there will be a degeneracy
$\Omega(\mu)$, giving the number of macrostates of measure $\curm(\mu)$.\footnote{In the case of
our examples $\Omega(\mu)$ will equal the number of distinct permutations of
$\{N_1,\ldots,N_p\}$.} We define
\begin{equation}
\Upsilon(\mu)=\frac{\curm(\mu)\Omega(\mu)}{\curm(\Lambda)},
\tripsep\mbox{where}\tripsep
\curm(\Lambda)=\sum_{\{\mu\}}\curm(\mu)\label{typ22}
\end{equation}
and $(S_\tB)_{\Upsilon}$ denotes the Boltzmann entropy for a macrostate
giving the largest value of $\Upsilon(\mu)$. With
\begin{equation}
S_\Lambda=S_\tB(\Lambda)= k_\tB\ln[\curm(\Lambda)],
\label{typ2bis3}
\end{equation}
the {\em entropy of} $\Lambda$,
it is clear that the inequalities $S_{\Lambda}>(S_\tB)_{\mx}\ge (S_\tB)_{\Upsilon}$
must be satisfied\footnote{Assuming that the Boltzmann entropy is
additively scaled so as to be non-negative.} and, from (\ref{typ4}),
{\mathindent=0cm
\begin{equation}
\frac{S_{\Lambda}-(S_\tB)_\mx}{Nk_\tB}\le \frac{S_{\Lambda}-(S_\tB)_\Upsilon}{Nk_\tB }<
- \frac{1}{N}\sum_{\{\mu\}} \frac{\curm(\mu)}{\curm(\Lambda)}\ln\left[\frac{\curm(\mu)}{\curm(\Lambda)}\right]
<
\frac{\ln[\eta(N,p)]}{N},
\label{extra3gis}
\end{equation}
where}
\begin{equation}
\eta(N,p)= \frac{(N+p-1)!}{N!(p-1)!}.
\label{extra3agis}
\end{equation}
%------------------------------------------------------------------------------------
\begin{figure}[t]
\begin{center}
\includegraphics[width=100mm,angle=0]{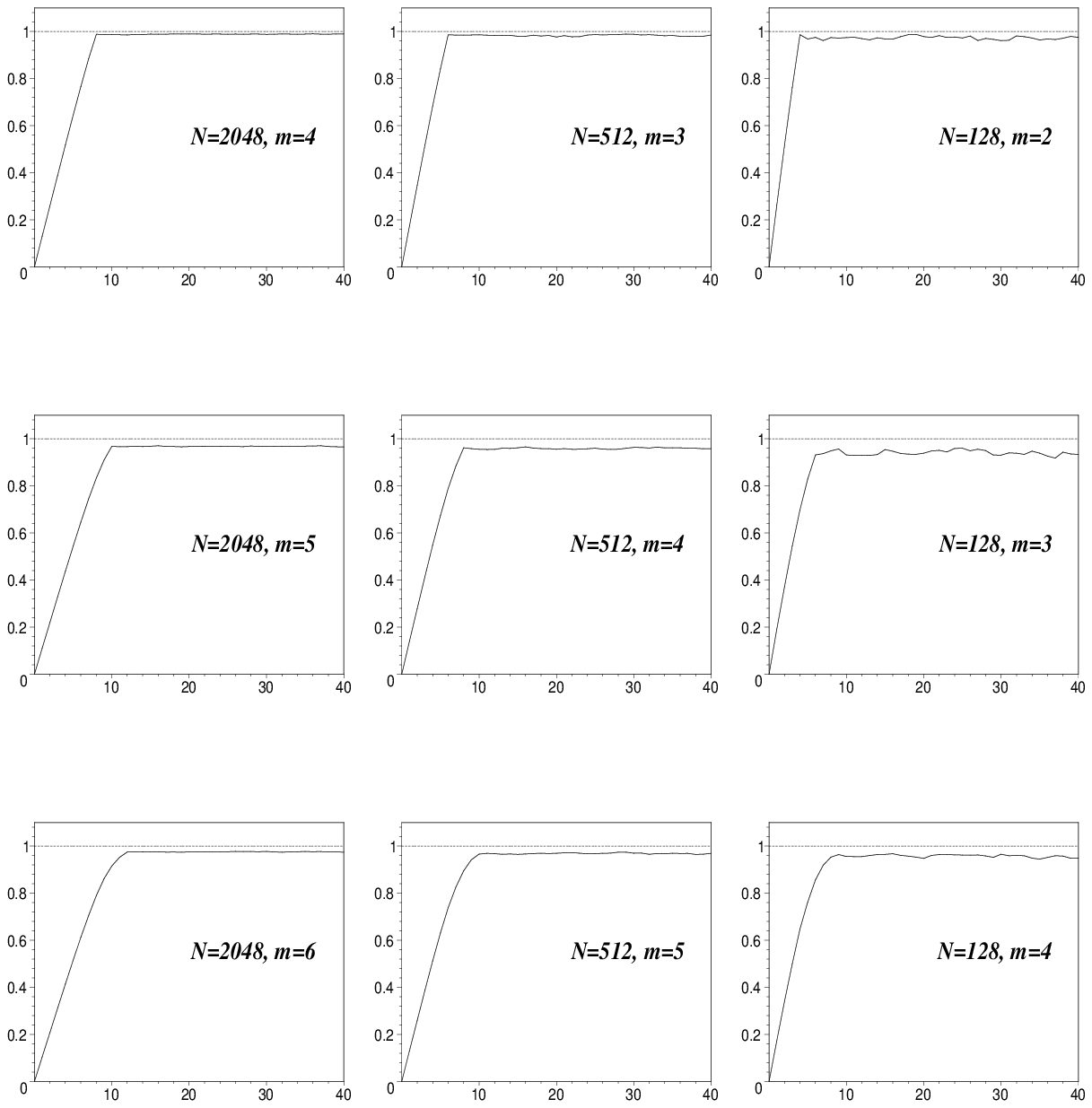}
\end{center}
\caption{The scaled Boltzmann entropy as it evolves with time for the baker's gas model.
Initially all the particle are randomly distributed in cell (1,1), ($N_1=N$).
The figures in horizonal lines correspond (from top to bottom) to average numbers of
particles per cell of 8, 2 and 0.5 respectively.
}\label{equ-fig4}
\end{figure}
%------------------------------------------------------------------------------------
\noindent In Fig.\ \ref{equ-fig4} we show the Boltzmann entropy of the baker's gas
for various cases of different numbers of particles and cells,
where initially the gas starts
with all the particles in cell
(1,1).\footnote{In all graphs of the Boltzmann entropy and Table \ref{tab1}, the minimum
entropy is set to zero with the condition $p=V$. Additionally in Fig.\ \ref{equ-fig4}
the entropy is scaled by its maximum value.}
We observe a rapid rise of the scaled entropy from its initial value
of zero followed by small fluctuations around a value rather less than the scaled value of
unity.\footnote{This behaviour is, of course, typical of a wide range of models with both
discrete and continuous phase spaces \citep{lavis3a}.}
It is clear that, in all the cases shown in Fig.\ \ref{equ-fig4}, $\Upsilon(\mu)$ is not maximal
for the largest macrostate, although detailed computations with large values of $N$ are not only difficult,
but not particularly revealing.

It is more useful to concentrate on small values of $N$ and $p$
where the results are easy to compute. With $p=4$ cells ($m=1$)
and $N=8$ we denote by the vector
$\bmf{n}=(n_1,n_2,n_3,n_4)$ an unordered distribution of particles in the four cells,
with $\Omega(\bmf{n})$ being the number of macrostates with this distribution.
Data for this case are given in order of decreasing
$\Upsilon(\bmf{n})$ in Table \ref{tab1}.
\begin{table}
\begin{tabular}{|c|c|c|c|c|}
\hline
$\bmf{n}$&$\Omega(\bmf{n})$&$\curm(\bmf{n})/
\curm(\Lambda)$&$S_\tB(\bmf{n})/(Nk_\tB)$&$\Upsilon(\bmf{n})$\\
\hline
\hline
(3,2,2,1)&12&0.0256&0.9283&0.3076\\
\hline
(4,2,1,1)&12&0.0128&0.8417&0.1538\\
\hline
(3,3,1,1)&6&0.01709&0.8776&0.1025\\
\hline
(3,3,2,0)&12&0.0085&0.7910&0.1025\\
\hline
(4,3,1,0)&24&0.0043&0.7043&0.1025\\
\hline
(4,2,2,0)&12&0.0064&0.7550&0.0769\\
\hline
(5,2,1,0)&24&0.0026&0.7271&0.0615\\
\hline
(2,2,2,2)&1&0.0385&0.9790&0.0385\\
\hline
(5,1,1,1)&4&0.0051&0.7271&0.0205\\
\hline
(5,3,0,0)&12&0.0009&0.5032&0.0103\\
\hline
(6,1,1,0)&12&0.0009&0.5032&0.0103\\
\hline
(4,4,0,0)&6&0.0011&0.5311&0.0064\\
\hline
(6,2,0,0)&12&0.0004&0.4165&0.0052\\
\hline
(7,1,0,0)&12&0.0001&0.2599&0.0014\\
\hline
(8,0,0,0)&4&$0.15\times10^{-4}$&0&$0.61\times10^{-4}$\\
\hline
\end{tabular}
\caption{Data for the baker's gas with $N=8$ and $p=4$.}\label{tab1}
\end{table}
\begin{figure}[t]
\begin{center}
\includegraphics[width=125mm,angle=0]{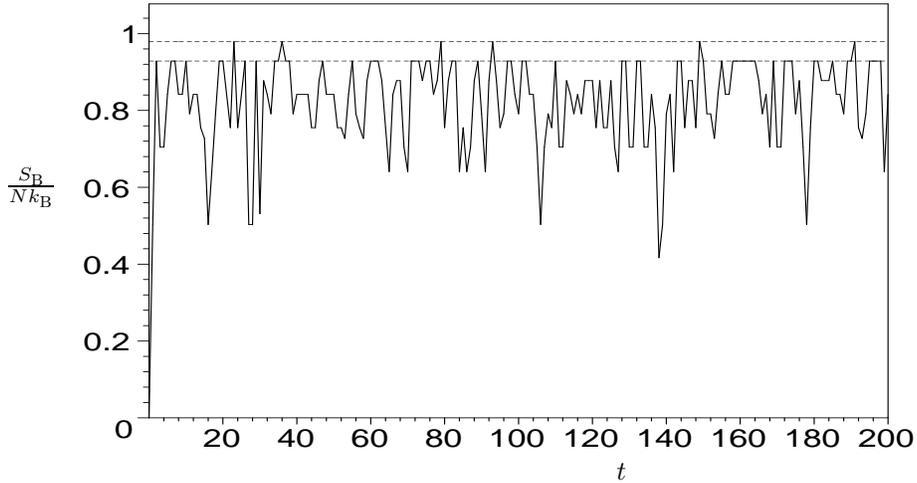}
\end{center}
\caption{The scaled Boltzmann entropy for the baker's gas model with $N=8$ particles
and $p=4$. Initially all the particle are randomly distributed in cell (1,1),
with configuration (8,0,0,0) and degeneracy of four.
Horizontal lines at $(S_\tB)_{\mx}$ and $(S_\tB)_\Upsilon$ are shown.}\label{equ-fig5}
\end{figure}
It will be observed that the macrostate of maximum
size $\bmf{n}=(2,2,2,2)$ is the eighth in the list. Values derived from Table \ref{tab1} are
$S_{\Lambda}=Nk_\tB\,1.3863$, $(S_\tB)_{\mx}= Nk_\tB\, 0.9790$ and
$(S_\tB)_{\Upsilon}=Nk_\tB\,0.9283$.
A typical evolution
of the entropy for this case is shown in Fig. \ref{equ-fig5}. The `plateau' at
$(S_\tB)_{\Upsilon}$ will be observed.

So if we were to propose a definition of the equilibrium macrostate what would we choose?
If we take a quasi-thermodynamic view and suppose that the equilibrium value for the
entropy is that into which it settles, with subsequent small fluctuations (and very rare large
fluctuations) then for small systems we have shown by simulation that this is not the largest entropy,
for which the system would be in the largest macrostate. An obvious strategy would be to
broaden the range of entropy for equilibrium. One might, for example, say that the system is
in equilibrium if $S_\tB\in [(S_\tB)_\mx,(S_\tB)_\mx(1-\varepsilon)]$,
where $\varepsilon= 2[(S_\tB)_\mx-(S_\tB)_\Upsilon]/(S_\tB)_\mx$.
A similar kind of `$\varepsilon$--equilibrium' strategy, in the context of the Gibbs approach, was proposed by
\citet{vanL2}. In our case, of course, this means that there will be small fluctuations
of entropy within equilibrium and the possibility of evolution
out of equilibrium accompanied by a large downward fluctuation in entropy. In addition to the
arbitrary division into macrostates, we have added a
demarcation given by the choice of value for $\varepsilon$. This is a simple demonstration of our contention
that the quality which we are trying to capture is a matter of degree,
rather than the two-valued property of either being {\em in equilibrium} or
{\em not in equilibrium}.
We, therefore, make the following proposal:
\begin{quote}
\textsl{All references to a system being, or not being,
in equilibrium should be replaced by references to
the commonness of the state of the system, with this property being given by some suitably-scaled
monotonically increasing function of the Boltzmann entropy.}
\end{quote}
Of course, at this stage, the proposal is confined to the Boltzmann approach. However, we hope
to show that its implementation is of relevance to the development of a view of statistical mechanics
which incorporated both the Boltzmann and Gibbs approaches and their connection to thermodynamics.
It should be emphasized that no connection is implied,
at this stage, between commonness and `commonness of occurrence'. We are not assuming that a
microstate with large commonness (a common microstate) belongs to a macrostate
which is visited more often by an evolving system. This connection is related to the question of {\em
typical behaviour.}
\subsection{Thermodynamic-Like Behaviour}\label{typical}
A favoured term of the Neo-Boltzmannians (but not of Boltzmann himself) is `typical'. Thus, in the
quote from \citet{leb1} in Sec.\ \ref{intro}, he refers to `typical microstates'
and `typical behaviour' and, in a later paper \citep{leb3}, he makes the
assertion that
\begin{quote}
``$S_\tB$ will {\em typically} increase in a way which
{\em explains} and describes qualitatively the evolution towards equilibrium of
macroscopic systems. Typical, as used here, means that the set of microstates corresponding to
a given macrostate [$\mu$] for which the evolution leads to a macroscopic decrease
in the Boltzmann entropy during some fixed time period $\tau$, occupies a subset of
[$\mu$] whose [measure] is a fraction of [$\curm(\mu)$] which goes very rapidly
(exponentially) to zero as the number of [particles] in the system
increases'' (p.\ S348).\footnote{We have used square brackets to replace Lebowitz's
notation by our own.}\label{lebpage}
\end{quote}
Now in ordinary usage `typical' is a description applied to an element of a set
with respect to some particular property.\footnote{Thus $a\in \mathcal{A}$ is typical
of $\mathcal{A}$ with respect to the property $b$ if $a$ and most members of
$\mathcal{A}$ have the property $b$. Or alternatively $b$ is typical of the set
$\mathcal{A}$ if a randomly chosen member of $\mathcal{A}$ is very likely to have
the property $b$.} In most instances of the use of `typical' by the Neo-Boltzmannians
the element of interest is a microstate (phase point) picked from phase space (or an
energy surface in phase space). The property
is `imminent evolution towards equilibrium indicated by an increase in Boltzmann entropy'
or perhaps more precisely `the absence of a imminent large decrease in Boltzmann entropy'.

 \citeauthor{leb3} ties this behaviour to the structure of macrostates.
This connection could be amplified in the following way. For the macrostate $\mu$ let
$\mu^{(++)}$ consist of all $\bx\in\mu$ such the the immediately next and immediately
previous macrostates along the trajectory are of measure larger than $\mu$.\footnote{For easy of
discussion we ignore the possibility of contiguous macrostates of equal measure.} With $\mu^{(--)}$
and $\mu^{(\pm\mp)}$
defined in a similar way, it follows, from the reversibility of the system and property (iv) of the macrostates
listed in Sec. \ref{approaches}, that
$\curm(\mu^{(-+)})=\curm(\mu^{(+-)})$.
The Neo-Boltzmannian argument is now based on the supposition
that
\begin{equation}
\curm(\mu^{(++)})\gg\curm(\mu^{(\pm\mp)})\gg\curm(\mu^{(--)}),
\label{extra104}
\end{equation}
for all (or most) macrostates $\mu$,
with the inequality becoming increasingly pronounced as $N$ increases.
Thus, for an arbitrarily chosen macrostate, the measure
$\curm(\mu^{(++)}\cup\mu^{(-+)})$ of phase points which lead to an increase in entropy
is very much larger than the measure $\curm(\mu^{(+-)}\cup\mu^{(--)})$
of phase points which lead to a decrease in entropy.
Then a typical (randomly chosen) point in $\mu$ belongs to $\mu^{(++)}\cup\mu^{(-+)}$,
leading to an increase in entropy. This is justified by \citet{bric1}
using Bayesian arguments. No explanation is needed for
the typical increase in entropy, beyond the application of the law of large numbers
in the limit of large $N$. If typical behaviour differs from this then an explanation
must be sought (maybe a re-examination
of the designation of macrostates). Within its own terms this Bayesian gloss of typicality
is perfectly satisfactory. However, it is predicated on the validity
of (\ref{extra104}). While this appears to be true for many toy models\footnote{It is easy
to verify for particular cases of the Kac ring model, which is the example most
often cited by \citeauthor{bric1}.} general rules for its validity are more difficult
to formulate.

But most statistical mechanical explanations
are concerned with something more than just the immediate increase in entropy
following the arbitrary selection of an initial phase point. The object of interest is the evolution along a
trajectory and the Bayesian argument does not adequately account for this.
Consider the following scenario.
For a system a phase point $\bx$ is chosen. It happens to be in the macrostate $\mu$ and according
to good Bayesian (or Laplacian) arguments and the inequality (\ref{extra104}) it will typically lie
in $\mu^{(++)}\cup\mu^{(-+)}$. Now suppose, in the course of evolution, the contiguous transition $\bx\to\bx'$
occurs to a macrostate $\mu'$. By definition $\curm(\mu')>\curm(\mu)$ and
$\bx'\in {\mu'}^{(-+)}\cup{\mu'}^{(--)}$. Now if $\bx'$ were chosen randomly in
${\mu'}^{(-+)}\cup{\mu'}^{(--)}$ we might typically expect, from (\ref{extra104}), that
$\bx'\in {\mu'}^{(-+)}$, leading to a further increase in entropy.
However, this is not necessarily the case. The dynamics determines the location
of $\bx'$ in ${\mu'}^{(-+)}\cup{\mu'}^{(--)}$.
Merely to associate probabilities with measure without random selection is not adequate
to close the argument.

The requirement of statistical mechanics is that {\em a typical trajectory is\break thermodynamic-like},
according
to the definition of this term on page \pageref{thermlk}. As yet this definition lacks precision.
However, it is certainly the case that if, along a trajectory and for most macrostates $\mu$,
the phase point spent an amount of time in $\mu$ proportionally related to
$\curm(\mu)$
the requirement of thermodynamic-like behaviour would be satisfied. To be more precise,
for any $\bx\in\Lambda$, let $\mcL_{\bx}$ be the trajectory through $\bx$,
and $\ttime_{\bx}(\mu)$ be the proportion of the time that the phase point on this trajectory is
in $\mu$.\footnote{It was shown by \citet{birk} that
$\ttime_{\bx}(\mu)$ exists and is independent of the location of $\bx$ on $\mcL_{\bx}$
for almost all $\bx\in\Lambda$; that is except possibly
for a set of $\curm$-measure zero. From this it follows \citep[see e.g.][]{lavis1}
that $\ttime_{\bx}(\mu)$ is a constant of motion almost everywhere in $\Lambda$.}
Now we can define {\em thermodynamic-like behaviour along the trajectory} $\mcL_{\bx}$ as the situation
where $\ttime_{\bx}(\mu)$ is approximately proportional to $\curm(\mu)$.
The more common a macrostate is the more common will be its occurrence along a trajectory.
For this behaviour to be {\em typical} it must be true for most trajectories.
This will hold in strong sense if
\begin{equation}
\ttime(\mu)=\tau(\mu),\hspace{1cm}\mbox{where}\hspace{1cm}\tau(\mu)=\curm(\mu)/\curm(\Lambda)
\label{prob2}
\end{equation}
for all $\mu$ and all trajectories apart from a set of measure zero. This is, of course,
the condition that the system is {\em ergodic}. So ergodicity is {\em sufficient}
for (perfect) thermodynamic-like  behaviour along all but an atypical set of trajectories of
measure zero.

The baker's transformation is a Kolmogorov system \citep[p.\ 91]{la&ma}
and is, therefore, both ergodic and mixing.\footnote{In fact
it is a Bernoulli system \citep[p.\ 125--126]{a&a}.}
So we have an explanation for the typical thermodynamic-like behaviour exhibited in
Figs.\ \ref{equ-fig4} and \ref{equ-fig5}. It is also not difficult to describe at least
some of the measure-zero atypical trajectories.  Suppose that the initial
binary strings giving the particle positions were periodic and the periods of all the particles were
commensurate. Then the behaviour of the system (detected, for example, by its
entropy profile) would have a periodic form at variance with thermodynamic-like
behaviour. The perfect gas of Sec.\ \ref{gas} is also ergodic (but not mixing)
as long as the particle speeds
are incommensurate \citep[p.\ 96--98]{farq}\footnote{The proof of this given by \citeauthor{farq},
which is a version of the discussion by \citet[p.\ 58--62]{khin}, is for two independent particles with
periodic boundary conditions and ergodicity is over the reduced manifold defined by the normal
integrals of the motion. Our model maps into one with periodic boundary conditions and particle velocities
to the right by `unfolding' the interval $[-L/2,L/2]$ into $[-L,L]$
\citep{lavis4}.
Then it can be seen that the only independent normal integrals of the motion are $\dot{x}^{(k)}=v^{(k)}$,
$k=1,2,\ldots,N$ and the motion is ergodic over the hypercube $x^{(k)}\in [-L,L]$, $k=1,2,\ldots,N$.}
and the atypical trajectories correspond to a choice of initial speeds which are commensurate and
produce periodic, rather than quasi-periodic behaviour.

However, ergodicity is not {\em necessary}\footnote{\citet{bric} also argues that ergodicity
is not sufficient, but this is based on the use of ergodicity to justify measurements as infinite time-averages.}
since (\ref{prob2}) is stronger than
is required for a typical system to behave in a thermodynamic-like
way and atypicality can also be broadened to include a set of trajectories of small but non-zero measure.
As pointed out by \citeauthor{bric}(\citeyear{bric}, \citeyear{bric1}) and verified by numerical calculations
\citep{lavis3a} the Kac ring model, which is not ergodic,\footnote{Its phase space divides into cycles.}
exhibits clear thermodynamic-like behaviour.

The typicality of thermodynamic-like behaviour, away from the perfect case of an ergodic system, is
difficult to formulate. What counts as sufficiently close to the perfect situation of (\ref{prob2}) to
yield an entropy profile which looks thermodynamic-like and how much atypical behaviour can be
tolerated? For the purpose of this work we shall leave these questions open with symbols of
approximation standing as markers for a more detailed investigation.\footnote{Quantification of
what counts as thermodynamic-like and then what count as a tolerable degree of atypical
non-thermodynamic-like behaviour could both be developed using the sort of $\varepsilon$ criteria described above for
$\varepsilon$-equilibrium.}

Ergodicity on $\Lambda$ is, of course, equivalent to $\Lambda$ being metrically indecomposable under the flow,
and a non-ergodic situation where we can at least map out the elements of the discussion
is when there is a finite or denumerable ergodic decomposition of $\Lambda$.
That is, to within subsets of
zero measure,
\begin{equation}
\Lambda=\displaystyle{\bigcup_{\{\kappa\}}}\, \Lambda_{\kappa}, \label{tl1}
\end{equation}
where each component $\Lambda_\kappa$ is invariant and metrically indecomposable under the flow.
Ergodic decomposition will, of course, apply to any model (like, for example, the Kac ring) where
$\Lambda$ consists of a finite number of discrete points; then each $\Lambda_\kappa$ is a
cycle of states. Ergodic decomposition also applies where the system is $\varepsilon$-ergodic \citep{vran}.
In this case one element of the decomposition $\Lambda_1$ (say) is such that
$\curm(\Lambda_1)=(1-\varepsilon)\curm(\Lambda)$ for small $\varepsilon>0$.\footnote{\citet{vran} has argued that
many dynamic systems of interest are $\varepsilon$-ergodic, but as yet the hard evidence for this
is lacking.}

When (\ref{tl1}) applies, the time spent in the macrostate $\mu$ for all
trajectories $\mcL_{\bx}$ with $\bx \in \Lambda_\kappa$
(apart from at set of measure zero) is
\begin{equation}
\ttime_\kappa(\mu)=\curm(\mu\cap\Lambda_\kappa)/\curm(\Lambda_\kappa).
\label{prob2a}
\end{equation}
For such a system it is clear that temporal
behaviour will be the same for all trajectories\footnote{At all relevant places in the following
discussion the phase: `except possibly a set of measure zero', is taken to apply.} within a $\Lambda_\kappa$.
The division must
be between those members of the ergodic decomposition in which the behaviour is thermodynamic-like
and those in which it is not. So:
\begin{enumerate}[T1)]
\item We indicate that the system is behaving in $\Lambda_\kappa$ in a way which can be counted as thermodynamic-like
by writing
\begin{equation}
\ttime_\kappa(\mu)\approxeq \tau(\mu),\hspace{1cm}\mbox{$\forall$  $\mu$.}
\label{prob2b}
\end{equation}
\item We denote by $\Lambda^{(\tT)}$ the union of all $\Lambda_\kappa$ satisfying (\ref{prob2b})
with $\Lambda^{(\tA)}= \Lambda\backslash\Lambda^{(\tT)}$ and by $\prbs_\kappa$ the probability\footnote{The meaning of which is
discussed in Sec. \ref{prob}.} that the system is in $\Lambda_\kappa$.
Then thermodynamic-like behaviour is typical for the system if
\begin{equation}
\sum_{\Lambda_\kappa\subset\Lambda^{(\tA)}} \prbs_\kappa\ll 1.
\label{prob2c}
\end{equation}
\end{enumerate}
%---------------------------------------------------
\section{The Gibbs Approach}\label{gibbs}
Given that we are concerned to incorporate both Boltzmannian and Gibbsian view of statistical mechanics
into our overall picture, we must now examine the role to be ascribed to the Gibbs approach.
This we do by means of a simple example.

Consider a gas of $N$ particles moving in $d$ dimensions.
Then the phase
vector $\bx\in\Gamma$ is $2dN$-dimensional, composed of configuration
and momentum vectors $\bq^{(k)}=(q^{(k)}_1,\ldots,q^{(k)}_d)$ and
$\bp^{(k)}=(p^{(k)}_1,\ldots,p^{(k)}_d)$, $k=1,2,\ldots,N$. Let $\Lambda$ correspond
to the gas being confined to the hypercubic box
$\mcB=\big\{\bx\big|-\tfrac{1}{2}L\le q^{(k)}_\alpha\le\tfrac{1}{2}L,\hspace{0.2cm}
\forall\,k,\forall\,\alpha\big\}$.
Then, if it is left undisturbed `a sufficient time to attain
thermodynamic equilibrium', the phase-point $\bx$ will, according to the Gibbs prescription,
be distributed in $\Lambda$ according to the  appropriated equilibrium
probability density function, which we denote by $\rho_\tG(\bx)$.

Now suppose that
the gas is confined by a partition to the part of the box, denoted by $\mcB^{(-)}$,
with $q_1^{(k)}<0$,
$\forall\,k$. In this situation, if the system is left to attain thermodynamic equilibrium,
the appropriate probability density function will be $\rho_\tG^{(-)}(\bx)$,
which differs from $\rho_\tG(\bx)$ only in respect of the restriction on the configuration
space. If the partition is removed at time $t=0$ the probability density function is
\begin{equation}
\rho(\bx;0)=\left\{\begin{array}{ll}
\rho_\tG^{(-)}(\bx),\pairsep &\bx\in\Lambda^{(-)},\\[0.3cm]
0,&\mbox{otherwise},
\end{array}\right.
\label{gibbs5}
\end{equation}
where $\Lambda^{(-)}$ is that part of $\Lambda$ corresponding to
all the particles being in $\mcB^{(-)}$. This is no longer the equilibrium distribution;
it will be the initial condition for a non-equilibrium
solution $\rho(\bx;t)$ evolving according to Liouville's equation. However, as we have indicated,
the Gibbs entropy for this solution remains constant and $\rho(\bx;t)$ does not converge to
$\rho_\tG(\bx)$ either in finite time, as we would like for thermodynamics, nor even
on an infinite time scale. The most we can obtain is, for a mixing system, when the expectation values
of a certain class of phase functions calculated using $\rho(\bx;t)$ converge, as $t\to\infty$,
to their expectation values calculated with $\rho_\tG(\bx)$.
The resolution to this problem suggested by \citet[p.\ 148]{gib}
\citep[see also][]{ehr2} was to coarse-grain the phase-space
in the manner in which macrostates were obtained in the Boltzmann approach.
In the macrostate $\mu$ the probability density function $\rho(\bx)$
is replaced by $\curpi(\mu)/\curm(\mu)$ where
\begin{equation}
\curpi(\mu)=\int_\mu \rho(\bx) \dd \curm,
\label{extra20}
\end{equation}
is the probability of the phase point $\bx$ being in $\mu$. Then
from (\ref{gibbs3})
\begin{equation}
(S_\tG)_{\tC\tG}[\curpi]= -k_\tB\sum_{\{\mu\}} \curpi(\mu)\ln[\curpi(\mu)]
+k_\tB\sum_{\{\mu\}} \curpi(\mu)\ln[\curm(\mu)].
\label{extra21}
\end{equation}
The objections to this approach are
well-known\footnote{See \citet{rid} for a recent
discussion
and \citet{rid&red} and \citet{lavis4} for the application of course-graining to the spin-echo system.}
and we shall not discuss them here. We shall suggest a different approach.

Suppose that a phase point $\bx_0\in\Lambda^{(-)}$.
What is the probability of the system being in a small measurable set $\curdelta\Lambda_0$ around $\bx_0$?
This will clearly depend on the physical circumstances of the system and will differ
according to whether the phase point is confined to $\Lambda^{(-)}$ by the physical partition.
In that case, if the system is in `thermodynamic equilibrium' the probability will be
$\rho_\tG^{(-)}(\bx_0)\curm(\curdelta\Lambda_0)$. However, if the partition is not present
and the system is in `thermodynamic equilibrium' the probability will be
$\rho_\tG(\bx_0)\curm(\curdelta\Lambda_0)$. Since $\bx_0$ corresponds to all the particles of
the gas being in one end of the box, it will (in our Boltzmann language) be a
rather `uncommon state' and we expect that
$\rho_\tG(\bx_0)\curm(\curdelta\Lambda_0)\ll\rho_\tG^{(-)}(\bx_0)\curm(\curdelta\Lambda_0)$.
Now consider the case where at time $t=0$ the partition is removed. According to the Gibbs prescription,
the probability of $\bx\in\curdelta\Lambda_0$ is $\rho(\bx_0;0)\curm(\curdelta\Lambda_0)
=\rho_\tG^{(-)}(\bx_0)\curm(\curdelta\Lambda_0)$.
The removal of the partition has not affected the probability. But compare this situation with that
where there has never been a partition present and the system phase point is in $\curdelta\Lambda_0$. What
physically distinguishes the two cases? If, in the course of its dynamic flow, the
system is at $\bx_0$ at $t=0$, its forward evolution will not affected by whether a partition
has just been removed or whether it simply happens to have evolved into this state. So why should
a different probability distribution be assigned to these two situations when
$t>0$?\footnote{Of course,
the trajectories for $t<0$ will differ and the system from which the partition was removed can be regarded
as suffering from `false memory syndrome'.} A consistent approach, consonant with our treatment of the
Boltzmann approach, is to suppose that the only meaningful probability density function to be used from
the Gibbs approach is the time-independent solution of Liouville's equation determined by the dynamics
and the physical constaints on the system. A change of physical constraints will lead to an instant
discontinuous change in the probability density function and the Gibbs entropy. An uncommon state
(like, for example, the case of all the particles being in one end of the box)
will have low probability when calculated using $\rho_\tG(\bx)$ and low Boltzmann entropy, but the same
Gibbs entropy as any other configuration.
%----------------------------------------------------------------------
\section{Probability}\label{prob}
In the Gibbs approach of Sec.\ \ref{gibbs} we introduced the probability density
function $\rho(\bx)$ on the invariant set $\Lambda$ and in terms of this, in (\ref{extra20}),
defined the probability $\curpi(\mu)$ that the phase point will be in the macrostate
$\mu$. What we have not done is define what we mean by probability.
Within the Gibbs approach this is most frequently done using ensembles, which
means that the probability density is not the property of a single system.
But we have already indicated, in Sec.\ \ref{intro}, that our object of interest is a single
system. Our aim in this work is to bring some kind of reconciliation between the Gibbs
and Boltzmann approaches and, as asserted by \citet{leb1} (see the quotation in Sec.\ \ref{intro}),
this latter neither has nor needs ensembles.

We need two probabilities, that the phase point is moving on a particular trajectory and that on that
trajectory the phase point lies in a particular subset of $\Lambda$. For the latter we shall
follow \citet{vonP2} in using the {\em time-average} definition
of probability. This analysis is simplified if, as we shall do, the ergodic
decomposition (\ref{tl1}) is assumed. Then:
\begin{enumerate}[(i)]
\item The probability $\prbs_\kappa=\prb(\bx\in\Lambda_\kappa)$ was introduced in Sec.\ \ref{typical}.
It can be taken to mean either the probability that when we choose the initial point $\bx(0)$
it is in $\Lambda_\kappa$ or the probability that when we investigate the system we find the phase point
in $\Lambda_\kappa$. In either case the probability is susceptible to
a Bayesian or Laplacian interpretation.
\item The conditional probability
that $\bx\in\mu$ given that $\bx\in\Lambda_\kappa$ is, from (\ref{prob2a}),
\begin{equation}
\prb(\bx\in\mu|\bx\in\Lambda_\kappa)
=\ttime_\kappa(\mu)=\frac{\curm(\mu\cap\Lambda_\kappa)}{\curm(\Lambda_\kappa)},
\label{prob7}
\end{equation}
where the first equality represents the time-average definition of probability.
\end{enumerate}
Then
{\mathindent=0cm
\begin{equation}
\prb(\bx\in\mu)=\sum_{\{\kappa\}} \prb(\bx\in\mu|\bx\in\Lambda_\kappa)\,\prbs_\kappa
=\langle\prb(\bx\in\mu|\bx\in\Lambda_\kappa)\rangle_{\tE\tD},
\label{prob7a}
\end{equation}
where} $\langle\cdot\rangle_{\tE\tD}$ denotes the expection value over the ergodic decomposition.

In each ergodic component there is a unique probability density function invariant under the flow
given by
\begin{equation}
\rho_\kappa(\bx)=\left\{\begin{array}{ll}
1/\curm(\Lambda_\kappa), &\mbox{$\bx\in\Lambda_\kappa,$}\\[0.25cm]
0,&\mbox{otherwise}
\end{array}\right.
\label{prob7b}
\end{equation}
and
\begin{equation}
\rho(\bx)=\sum_{\{\kappa\}} \rho_\kappa(\bx)\,\prbs_\kappa=\langle \rho_\kappa(\bx)\rangle_{\tE\tD}.
\label{prob7c}
\end{equation}
Then, from (\ref{extra20}), (\ref{prob7b}) and (\ref{prob7c}),
\begin{equation}
\curpi(\mu)=\prb(\bx\in\mu)=\sum_{\{\kappa\}} \frac{\curm(\mu\cap\Lambda_\kappa)}{\curm(\Lambda_\kappa)}\,\prbs_\kappa,
\label{prob7ca}
\end{equation}
as given by
(\ref{prob7}) and (\ref{prob7a}).

For any functions $f$, integrable on $\Lambda$,
and $\eusG$, summable on the macrostates, the expectation
value, of $f$ with respect to $\rho$ and $\eusG$ with respect to $\curpi$, are respectively
\begin{equation}
\langle f\rangle_\rho= \int_\Lambda \rho(\bx) f(\bx) \dd\curm,\pairsep
\langle \eusG\rangle_\curpi=\sum_{\{\mu\}}\curpi(\mu)\eusG(\mu).
\label{prob101}
\end{equation}
Clearly
\begin{equation}
\langle f\rangle_\rho=\langle\langle f|\mu\rangle\rangle_\curpi,\tripsep\mbox{where}\tripsep
\langle f|\mu\rangle= \frac{1}{\curpi(\mu)}\int_\mu \rho(\bx) f(\bx) \dd\curm.
\label{prob102}
\end{equation}
The time-average of $\eusG$ along a trajectory in $\Lambda_\kappa$ is
\begin{equation}
\stackrel{\kappa}{\overline{\eusG(\mu)}}= \sum_{\{\mu\}}\ttime_\kappa(\mu)\eusG(\mu)
\label{prob6a}
\end{equation}
and, from (\ref{prob7}), (\ref{prob7a}) and (\ref{prob7ca}),
\begin{equation}
\big\langle\stackrel{\kappa}{\overline{\eusG(\mu)}}\big\rangle_{\tE\tD}=
\sum_{\{\kappa\}} \frac{\prbs_\kappa}{\curm(\Lambda_\kappa)} \sum_{\{\mu\}}
\curm(\mu\cap\Lambda_\kappa)\eusG(\mu)=
\langle \eusG\rangle_{\curpi}.
\label{prob6b}
\end{equation}
When the system behaviour in $\Lambda_\kappa$ is thermodynamic-like it follows, from (\ref{prob2}) and (\ref{prob2b}),
that
\begin{equation}
\stackrel{\kappa}{\overline{\eusG(\mu)}}\approxeq\sum_{\{\mu\}} \frac{\curm(\mu)}{\curm(\Lambda)} \eusG(\mu).
\label{prob6bb}
\end{equation}
A phase function $f$ which happens to be constant over each of the macrostates $\mu$
has $f(\bx)=\langle f|\mu\rangle$, $\forall\,\bx\in\mu$, giving
$\langle f\rangle_\rho=\langle f\rangle_{\curpi}$. In particular, from (\ref{prob7ca}),
\begin{equation}
\langle S_\tB\rangle_{\rho}=\langle S_\tB\rangle_{\curpi}
= k_\tB \sum_{\{\kappa\}} \frac{\prbs_\kappa}{\curm(\Lambda_\kappa)} \sum_{\{\mu\}}
\curm(\mu\cap\Lambda_\kappa)\ln[\curm(\mu)].
\label{prob5}
\end{equation}
Relating the Gibbs entropy to the Boltzmann entropy will necessitate some rational choice for the
probabilities $\{\prbs_\kappa\}$. One view of $\prbs_\kappa$ is that it is the probability
that, if we make a random choice for the the initial system point $\bx(0)$, then it lies in $\Lambda_\kappa$.
If we assume that all points of $\Lambda$ are equally likely then on Bayesian/Laplacean grounds and
consonant with the approach of \citet{bric1} we should choose
\begin{equation}
\prbs_\kappa=\curm(\Lambda_\kappa)/\curm(\Lambda).
\label{assum1}
\end{equation}
Then, from (\ref{prob7b}), (\ref{prob7c}) and (\ref{prob7ca})
\begin{equation}
\rho(\bx)=1/\curm(\Lambda), \pairsep \curpi(\mu)=\curm(\mu)/\curm(\Lambda),
\label{assum2}
\end{equation}
which is the microcanonical distribution. From
(\ref{gibbs3}) and (\ref{typ2bis3}),
\begin{equation}
S_\tG= S_{\Lambda}=k_\tB\ln[\curm(\Lambda)]
\label{mer1}
\end{equation}
and, from (\ref{prob5}),
\begin{equation}
\langle S_\tB\rangle_{\rho}=\langle S_\tB\rangle_{\curpi}=S_\tG+
k_\tB \sum_{\{\mu\}}\curpi(\mu)
\ln[\curpi(\mu)].
\label{mer1b}
\end{equation}
When the system behaviour in $\Lambda_\kappa$ is thermodynamic-like,
it follows, from (\ref{prob6b}) and (\ref{prob6bb}), that
\begin{equation}
\stackrel{\kappa}{\overline{\eusG(\mu)}}\approxeq
\big\langle\stackrel{\kappa}{\overline{\eusG(\mu)}}\big\rangle_{\tE\tD}=
\langle \eusG\rangle_{\curpi}
\label{mer1a}
\end{equation}
and, from (\ref{prob2c}), thermodynamic-like behaviour is typical if
\begin{equation}
\curm\left(\bigcup_{\Lambda_\kappa\subset\Lambda^{(\tA)}} \Lambda_\kappa\right)\ll \curm(\Lambda).
\label{prob2cbis}
\end{equation}
%--------------------------------------------------------------------------------------
\section{Proposals}\label{prop}
In this section we shall make the following assumptions:
\begin{enumerate}[(a)]
\item That the system has the ergodic decomposition
(\ref{tl1}).\footnote{Of course, the analysis includes the ergodic case where the decomposition
has one member with $\prbs_1=1$.}
\item That the probabilities $\{\prbs_\kappa\}$ for the system phase point being in members of
the decomposition are given by (\ref{assum1}).
\item That (\ref{prob2cbis}) applies and thus that thermodynamic-like behaviour is typical.
\end{enumerate}
Assumption (a) is a purely dynamic assumption which may or may not be true for any particular system.
The statistical assumption (b) is again based on the dynamics and has the same status as the assumption
one might make about a die being unbiased when making a sequence of trials.
Assumption (c) has a dynamic aspect but is also dependent
on some quantification of whether behaviour is or is not thermodynamic-like.

As indicated in Sec.\ \ref{intro}, two of the obstacles
to overcome, in reconciling the Boltzmann and Gibbs approaches,
are the different definitions of equilibrium
and of entropy. With regard to equilibrium we propose, as indicated in Sec.\ \ref{ttde}, that
the two-valued condition of the system being or not being in equilibrium
is replaced by a continuous property called {\em commonness}, based on a designation
of macrostates related to a set of macrovariables. A microstate is more or less common
according to whether the macrostate in which it is situated is of greater or smaller size.
The Boltzmann entropy is a measure of the size of macrostates and thus provides
a measure of commonness.
Gibbs methods with a time-independent probability density function are to be retained
as the practical means for obtaining an `analogue' of thermodynamics. The reason
for using a time-independent probability density
function to calculate thermodynamic properties is not that the system is in equilibrium but that
the underlying dynamic is autonomous.\footnote{A non-autonomous dynamic system would not yield
a time-independent solution to Liouville's equation.}

The question of entropy can be subsumed into a more general account of the relationship between
statistical mechanical and thermodynamic variables.
As we have indicated in Sec.\ \ref{approaches}, the standard Gibbs perception of the relationship
between statistical mechanical and thermodynamic variables is that they fall into three classes.
Those for which
\begin{enumerate}[C1)]
\item the corresponding statistical mechanical variable is related to a phase function.
\item the statistical mechanical variable and thermodynamic variable
are identical and equal to an external parameter of the system.
\item the corresponding statistical mechanical variable
is a functional of the probability density
function.
\end{enumerate}
 The primary example of C3 is the entropy given by the Gibbs formula
(\ref{gibbs3}).\footnote{The class into which a variable falls depends, of course, on the
environmental conditions of the system. Thus for a thermally open system the temperature
is an independent parameter, whereas for a thermally closed system (like the isolated
systems considered here) the temperature is a functional of $\rho$.}
In the interests of producing
a smooth relationship between the Boltzmann and Gibbs approaches we shall make two modifications
to the C1--C3 scheme. The first, which is part of the standard Boltzmann approach, is to related
thermodynamic entropy
to a phase function (which is also a macrovariable),
namely the Boltzmann entropy $S_\tB(\bx)$; so entropy is now of type C1.
The second is to propose a particular
relationship, in the case of C1 variables, between thermodynamic variables and phase functions.

In Sec.\ \ref{approaches} we introduced the set $\Xi$ of macrovariables defined on the macrostates $\{\mu\}$.
The scheme we now propose is that a thermodynamic variable, $F$ in class C1, is related to a phase
function $f(\bx)$, defined on $\Lambda$, via a macroscopic function $\eusF(\mu)\in\Xi$.
The statistical mechanical variable is the macrovariable $\eusF$.
In the case of entropy the phase function and macrovariable
are identical and equal to the Boltzmann entropy. For other thermodynamic quantities
this is not the case. We are, thus, proposing the three-part
scheme $f\rightsquigarrow\eusF\rightsquigarrow F$, where we need to explicate the relationships
denoted by `$\rightsquigarrow$'.
\begin{enumerate}[]
\item\underline{${\eusF\rightsquigarrow F}$}\,:
The value $\eusF(\mu)$ is the result of a measurement of $f$, course-grained,
to effectively give the same value throughout the macrostate $\mu$. This perception
together with the definition of macrostates and the Boltzmann entropy
in Sec.\ \ref{approaches} serves as the demarcation between the
microscopic and macroscopic realms. Of course,
this demarcation is to some extent arbitrary, but it is equally so for any
macroscopic physical theory.\footnote{See e.g. the definition of fluid
density in \citet[p.\ 1]{lan&lif1}.}
The value of the thermodynamic quantity
$F$, along the trajectory $\mcL_{\bx_0}$ passing through $\bx_0$,
we take to be equal to the average of the result of a large number of
measurements of $\eusF$ taken at arbitrarily chosen times.
So we can effectively define
\begin{equation}
F= \sum_{\{\mu\}} \ttime_{\bx_0}(\mu)\eusF(\mu).
\label{merger1}
\end{equation}
$F$ is a constant of motion, but not
constant on $\Lambda$ except when the system is ergodic. When $\Lambda$ has the ergodic
decomposition (\ref{tl1}), $F$ has the set of values
\begin{equation}
F_\kappa=\sum_{\{\mu\}} \ttime_\kappa(\mu)\eusF(\mu)=\,\,\stackrel{\kappa}{\overline{\eusF(\mu)}}.
\label{merger1a}
\end{equation}
From (\ref{mer1a}), when the behaviour in $\Lambda_\kappa$ is thermodynamic-like,
\begin{equation}
F_\kappa\approxeq\langle\eusF\rangle_\curpi.
\label{merger2aa}
\end{equation}
Measurement of the thermodynamic variable $F$ will typically lead to a value close to $\langle\eusF\rangle_\curpi$.

\item\underline{$f\rightsquigarrow\eusF$}\,: We now need to be more precise about
what we mean by the macrovariable $\eusF$. Ideally, of course, the identification
\begin{equation}
\eusF(\mu)=\langle f|\mu\rangle,
\label{merger3}
\end{equation}
in the notation of (\ref{prob102}), would be desirable, because then it follows
from (\ref{merger2aa}), that, for thermodynamic-like behaviour,
\begin{equation}
F_\kappa\approxeq\langle f\rangle_\rho.
\label{merger2ab}
\end{equation}
This, to within the tolerance indicated by `$\approxeq$', is the standard
identification between thermodynamic variables
and the expectation values of phase functions in Gibbs theory. As we have observed in Sec.\ \ref{prob},
(\ref{merger3}) is exactly true for the Boltzmann entropy because it is, by definition, constant
over each macrostate.
We should not expect this to be true for all phase functions. But, of course,
not all phase functions have a correspondence with thermodynamic variables. In particular
we shall assume\footnote{These are two of the properties assumed for a {\em finite-range observable}
function by \citet[p.\ 2--3]{lan1}, but, as he says, his use of the word `observable' is not intended
to have ``any profound significance''.} $f(\bx)$ is
(a) continuous on each macrostate and (b) invariant under permutations of the $N$ microsystems.
We have already assumed, in Sec.\ \ref{approaches} condition (iii),
that each macrostate is invariant under permutations of the $N$ microsystems. So the macrostate
$\mu$ can be divided into $n_\mu$ non-overlapping subsets $\lambda^{(j)}_\mu$, $j=1,2,\ldots,n_\mu$
each of identical $\curm$--measure. Representing by $\lambda_\mu$ a generic member of this set, it follows,
from (\ref{prob102}), that
\begin{equation}
\langle f|\mu\rangle=\frac{1}{\curpi(\lambda_\mu)}\int_{\lambda_\mu} \rho(\bx) f(\bx) \dd\curm.
\label{merger4}
\end{equation}
When macrostates are constructed from cells in the one-microsystem phase space,
 $n_\mu$ is the combinatorial factor in (\ref{typ4}) and
$\curm(\lambda_\mu)=\curm(\Lambda)/p^\tmN$. This latter becomes small as $p$ becomes large.
In this case, since $f(\bx)$ is continuous over $\Lambda(\mu)$, it is likely to vary very
little.\footnote{One might at this stage, augment the conditions on $f$, given above,
to the full set defining a finite-range observable (\citeauthor{lan1}, ibid). Then, for
large $N$, the results of \citeauthor{lan1} and \citet[p.\ 62--69]{khin} could
be used to assert that the dispersion of $f$ around $\langle f\rangle_\rho$ is small.}
Although the macrostate $\mu$ may be a very large part of $\Lambda$ it will be made of a
large number of cells in each of which $f(\bx)$ has the same nearly constant value.
Thus\footnote{The symbol `$\approx$' used here to indicated approximate equality, should be
distinguished from `$\approxeq$' which is used  in this paper with the special sense of
`approximately equal in a system for which thermodynamic-like behaviour is typical'.}
\begin{equation}
\eusF(\mu)\approx f(\bx),\pairsep \mbox{for any point $\bx\in\mu$}
\label{merger5}
\end{equation}
and
\begin{equation}
\langle\eusF\rangle_\curpi\approx\langle f\rangle_\rho,\pairsep
 \mbox{as the size $\curm(\lambda_\mu)$ becomes small for all $\mu$.}
\label{merger6}
\end{equation}
\end{enumerate}
So to recapitulate the steps of the argument:
\begin{enumerate}[(i)]
\item Thermodynamic variables are time-averages of phase functions $f$ course-grained (to produces
statistical mechanical variables) over macrostates.
\item The thermodynamic variable is approximated along (typical) thermodynamic-like trajectories by
the average of $f$ with respect to the probability density function $\rho$.
\end{enumerate}
This establishes the relationship between a Boltzmannian view of thermodynamic variables as time averages
and the Gibbsian view in which they are phase averages with respect to the appropriate probability
density function.

We now consider in more detail the case of entropy.
It is often said that in ``equilibrium [the Gibbs entropy] agrees with Boltzmann and Clausius entropies
(up to terms that are negligible when the number of particles is large) and everything is fine'' \citep[p.\ 188]{bric}.
Apart from the caveats we have entered in Sec.\ \ref{ttde} with respect to the concept of equilibrium,
this assertion is based on two propositions (i) that the appropriate probability density function for
the Gibbs approach is given by the microcanonical distribution (\ref{assum2})
and (ii) that for large $N$ the `equilibrium' macrostate effectively fills the whole of $\Lambda$.
The first of these we have established, within our approach, for a system with an ergodic decomposition
and with the choice of probabilities (\ref{assum1}).
In the case where macrostates are defined using the cell method described in Sec. \ref{bmod},
(ii) is not difficult to establish.
Using Stirling's formula with (\ref{extra3agis}),
{\mathindent=0cm
\begin{equation}
\frac{\ln[\eta(N,p)]}{N}\simeq\left\{
\begin{array}{ll}
\displaystyle\frac{(p-1)\ln(N)}{N},&\mbox{with fixed $p$,}\\[0.5cm]
(1+\xi)\ln(1+\xi)-\xi\ln(\xi),
&\mbox{with $p=\xi N$,}
\end{array}\right\}\hspace{0.15cm}\mbox{as $N\to\infty$.}
\label{extra33}
\end{equation}
So} in the case where $p$, the number of cells is fixed,
it follows from (\ref{extra3gis}) that
{\mathindent=0cm
\begin{equation}
(S_\tB)_\mx/N\rightarrow S_\tG/N, \tripsep (S_\tB)_\Upsilon/N\rightarrow S_\tG/N,\tripsep
\mbox{as $N\rightarrow\infty$.}
\label{extra33a}
\end{equation}
This} is in agreement with the results shown in the diagonal figures in Fig.\ \ref{equ-fig4}
where we see that the value at which the entropy settles down is closer to the maximum entropy
for larger values of $N$. In the second case $1/\xi$ is the number of microsystems per cell. With
both $N$ and $p$ large the convergence is effected by increasing the number of microsystems per
cell ($\xi\to 0$). This is illustrated by the figures in the vertical columns of
Fig.\ \ref{equ-fig4}.
\begin{figure}[t]
\begin{center}
\includegraphics[width=100mm,angle=0]{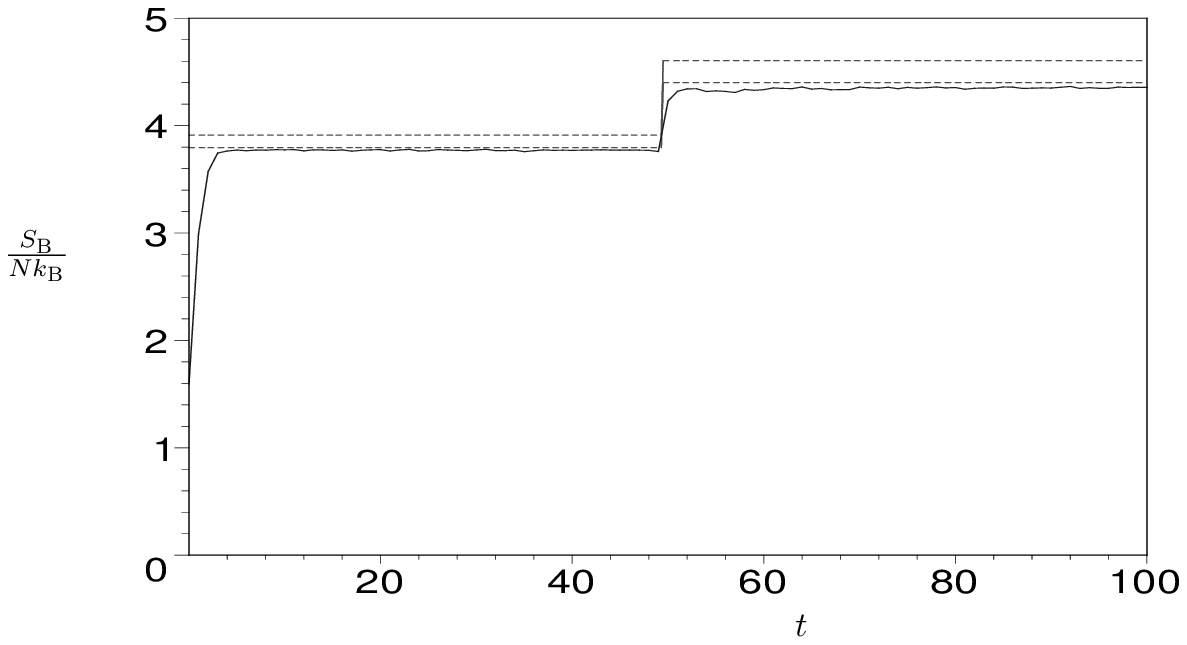}
\end{center}
\caption{The scaled Boltzmann entropy for the perfect gas model with $N=1000$, $p=100$.
The gas is allowed to evolved in the time interval $0\le t\le 50$ confined to the left
half of the box and then in the interval $50\le t\le 100$ into the whole box.
The upper and lower horizontal lines in each range correspond to $S_\tG$
and $(S_\tB)_{\mx}$ respectively.
}\label{equ-fig6}
\end{figure}
It also follows from (\ref{mer1a}), (\ref{mer1b}) and (\ref{extra3gis}) that
\begin{equation}
\stackrel{\kappa}{\overline{S_\tB(\mu)}}/N\,\approxeq\,\langle S_\tB\rangle_\rho/N\rightarrow S_\tG/N,\tripsep
\mbox{as $N\rightarrow\infty$.}
\label{extra33ab}
\end{equation}
As an illustration of these results we take the perfect gas
and suppose that it is first confined
by a partition to the region $x<0$. The Gibbs entropy,
given by (\ref{typ2bis3}) and (\ref{mer1}), is
$S_\tG^{(-)}=k_\tB N\ln(L/2)$ and $S_\tG=k_\tB N\ln(L),$
for the respective cases where the gas is confined by the partition
and where it is free to evolve over the whole box. The Boltzmann entropy
in the latter case is given by (\ref{typ2}) and (\ref{typ4}), with $V=L$
and in the former case by the same formula\, but restricted by the condition that
$N_\ell=0$ for $\ell>p/2$.\footnote{This expression is, of course, exactly the same
as that for a box of length $L/2$  partitioned into $p/2$ strips.}
We consider the situation in which the system starts in a minimum entropy macrostate
(with all the particles in the first strip). It is allowed to evolve over
the time interval $[0,50]$ in the region $x\le 0$. The partition is then
removed and it is allowed to evolve in the whole box in the time
interval $[50,100]$. This is shown in Fig.\ \ref{equ-fig6} where
$(S_\tG-(S_\tB)_\mx)/(Nk_\tB)$ in each range is well within the upper
bound given by (\ref{extra3gis}) and (\ref{extra33}).
%-------------------------------------------------------------------------------------------
\section{Conclusions}\label{conc}
In this programme we have used ergodicity and ergodic decomposition and as indicated
in Sec.\ \ref{intro} there is deep (and justified) suspicion of the use of ergodic
arguments to support the foundations of statistical mechanics, particularly
among philosophers of physics. Having given a comprehensive review of the problems
of ergodic arguments, \citet[p.\ 75]{ear&red} offer the opinion ``that ergodic
theory in its traditional form is unlikely to play more than a cameo role in
whatever the final explanation of the success of equilibrium
statistical mechanics turns out to be''.
In its `traditional form' the ergodic argument goes something
like this: (a) Measurement processes on thermodynamic systems take a long time compared to
the time for microscopic processes in the system and thus can be effectively regarded as
infinite time averages. (b) In an ergodic system the infinite time average can be shown, for
all but a set of measure zero, to be equal to the macrostate average with respect to an
invariant normalized measure
which is unique.\footnote{In the sense that it is the only invariant normalized
measure absolutely continuous
with respect to the Lebesque measure.}
The traditional objections to this argument are also well known: (i) Measurements
may be regarded as time averages, but they are not {\em infinite} time averages.
(If they were one could not, by measurement, investigate a system not in equilibrium.
In fact, traditional ergodic theory does not distinguish between systems in equilibrium
and not in equilibrium.) (ii) Ergodic results are all to within sets of measure zero
and one cannot equate such sets with events with zero probability of occurrence.
(iii) Rather few systems have been shown to be ergodic. So one must look for a reason for
the success of equilibrium statistical mechanics for non-ergodic systems and when it is
found it will make the ergodicity of ergodic systems irrelevant as well.

Our use of ergodicity differs substantially from that described above and it thus escapes
wholly or partly the strictures applied to it. In respect of the question of
equilibrium/non-equilibrium we argue that the reason this does not arise in ergodic arguments
is that equilibrium does not exist. The phase point of the system, in its passage along
a trajectory, passes through common (high entropy) and uncommon (low entropy) macrostates {\em and that is all}.
So, although in our definition of a thermodynamic variable we have extended a large finite
number of measurements to an infinite set of measurements, we cannot be charged with
`blurring out' the period when the system was not in equilibrium. The charge against
ergodic arguments related to sets of measure zero is applicable only if one wants
to argue that the procedure always works; that is that non-thermodynamic-like behaviour
never occurs. But we have, in this respect taken a Boltzmann view. We need thermodynamic-like
behaviour to be typical, but we admit the possibility of atypical behaviour occurring with small but
not-vanishing probability.\footnote{For
our model systems we have indicated in Sec.\ \ref{typical} the way in which atypical
behaviour can occur and \citet{lavis3a} demonstrates simulations of cases of this kind.}
While the class of systems admitting a finite or denumerable ergodic decomposition is likely
to be much larger than that of the purely ergodic systems,\footnote{As indicated above it will include any
system with a finite number of phase points.} there remains the difficult question of determining the conditions
under which the temporal behaviour along a trajectory, measured in terms of visiting-times
in macrostates, approximates, in most members of the ergodic decomposition, to
something recognizable as thermodynamic-like behaviour.
\section*{Acknowledgments}
I am grateful to Jos Uffink and Dennis Dieks for inviting me to make this contribution
to The ESF Conference
on Philosophical and Foundational Issues in Statistical Physics, Utrecht, November 2003.
I should also like to thank Harvey Brown, Michael Mackey, Gerard Watts and Ivan Wilde for useful discussions
and Janneke van Lith and the referees for their constructive comments on an earlier version
of the paper.

%\bibliography{/work/bibliography/references}
\end{document}